\def \vc #1{{\mbox{\boldmath $#1$}}}
\def\thetag{{\vc \theta}}
\def\alphag{{\vc \alpha}}
\def\nablag{{\vc \nabla}}
\def\boldg{{\bf g}}
\def\gammag{{\vc \gamma}}

\def\epsilong{{\vc \epsilon}}

\def\deltag{{\vc \delta}}
\def\xg{{\vc x}}
\documentstyle[editedvolume,psfig]{crckapb} 

\begin{opening}
\title{COSMOLOGICAL APPLICATIONS OF GRAVITATIONAL LENSING
       }

\author{Y. MELLIER}
\institute{Institut d'Astrophysique de Paris \\
           and Observatoire de Paris, DEMIRM\\
           98 bis Boulevard Arago\\
           75014 Paris \\
           France}

\end{opening}

\runningtitle{COSMOLOGICAL APPLICATIONS OF GRAVITATIONAL LENSING
}

\begin{document}


\section{Introduction}
Because gravitational lensing provides a unique tool to probe 
almost directly the dark matter, its use for cosmology 
 is of considerable interest. 
The discovery of giant arc(let)s in
clusters of galaxies (Soucail et al 1987, Lynds \& Petrosian 1986, 
 Fort et al 1988) 
 or Einstein rings around galaxies (Hewitt et al 1988), 
and the spectroscopic proofs that they
are produced by gravitational lensing effects 
 (Soucail et al 1988)
  have revealed that gravitational distortion 
 can probe with a remarkable amount of details
  the mass distribution of clusters
 (Tyson et al 1990, Kaiser \& Squires 1993) 
and galaxies (Kochanek 1990).
 \\
The applications of gravitational lensing to cosmology are so important
that one cannot ignore them in a course dedicated to observational
cosmology. The most obvious applications are 
the determination of the mass of gravitational systems, because 
 the total mass inferred from a simple gravitational lens model is 
 remarkably robust.  In this respect microlensing experiments as well 
 as strong and weak lensing studies provide the most powerful 
 techniques to probe the dark matter of the Universe from 
 the Jupiter-like planets up to large-scale structures. That is 
 more than 16 orders of magnitude in size, more than 19 orders
 of magnitude in mass, and more than 25 orders of magnitude in density
contrast!  These limits are only due to technical 
 limitations of present-day instrumentation, and in principle 
 gravitational lensing can probe a much broader range.  Therefore, 
 measuring the mass distribution with gravitational lensing 
 can put important constraints on the gravitational history 
 of our Universe and the formation of its structures and 
its virialized gravitational systems. \\
Gravitational
lensing also can provide valuable constraints on the cosmological
parameters. For examples, the 
 fraction of quasars with multiple-image, or  the magnification
bias in lensing-clusters depend on the curvature of the Universe.
  Deep surveys devoted to these lenses have already put boundaries 
on the cosmological constant $\lambda$. Furthermore, the
measurement of time delays of transient events observed in multiple
images of lensed sources can potentially provide useful constraints 
 on the Hubble constant, $H_0$, on cosmological scales where  
 distortion of the measurements from any local perturbation is negligible.
   Of equal
importance,
gravitational lenses can also be used as  gravitational telescopes in
order to
observe the deep Universe. When strongly magnified, detailed
structures of
extremely high redshift galaxies can be analyzed spectroscopically in
order to understand the dynamical stage and 
  the merging history  of the young distant
galaxies
(Soifer et al 1998) . Furthermore, the joint analysis of the dark matter
distribution
in clusters and the shape of the lensed galaxies can be used to recover
their redshift
distribution.\\
In this course, I will focus on some 
applications to cosmology. However, in order to avoid self-duplications,  
   I will only address some aspects about the mass determination from 
gravitational lensing studies. The other topics    
 more detailed presentations (namely, distant galaxies, 
cosmological parameters, lensing on the CMB) 
 or more detailed presentations can be found in 
  Bernardeau (this proceedings), Blandford \& Narayan (1992), 
 Fort \& Mellier (1994),  Mellier (1998, 1999),
 Refsdal \& Surdej (1994) and obviously in the textbook  written  
 by Schneider, Ehlers \& Falco (1992). The microlensing experiments will be also presented   
elsewhere (Rich, this proceedings). Since F. Bernardeau has already 
 discussed the theoretical aspects in his lecture, I will only make some 
addenda on some 
specific quantities in order to clarify the order of magnitude, but I 
assume that all the basic concepts are know already.  

\section{Some important quantities and properties}

\subsection{Image multiplicity and Einstein radius}
Let us assume the lens
 being  a point mass of mass $M$ and a
one-dimensional configuration. In that case, the
solutions
for $\theta_I$ are given by
\begin{equation}
\theta_S=\theta_I+{D_{LS} \over D_{OL} D_{OS}} {4 G M \over c^2
\theta_I}
\end{equation}
\noindent which in general has 2 solutions. This illustrates the fact that
gravitational
lenses can produce multiple images. In the special case of a point mass,
the
divergence of the deviation angle at the origin is meaningless. For a
more realistic
mass density, the divergence vanishes and this produces another solution
for the
images close to the center. This reflects a more general theorem that
the number
 of
lensed images is odd for non-singular mass density (Burke 1981). A
 critical case occurs when $\theta_S=0$, that is
when there is a perfect alignment between the source, the lens and the
observer.
The positions of the images are degenerated and form a strongly
magnified circle
at the {\sl Einstein radius} $\theta_E$:
\begin{equation}
\theta_E=\left[{4 G M \over c^2} {D_{LS} \over D_{OL}
D_{OS}}\right]^{1/2} .
\end{equation}
\noindent For example,
\begin{itemize}
\item for a star of 1 $M_{\odot}$ at distance $D$=1 kpc,
$\theta_E=0.001$ arcsec,
\item for a galaxy of 10$^{12}$ $M_{\odot}$ at $D$=1 Gpc, $\theta_E=1$
arcsec,
\item and for a clusters of 10$^{14}$ $M_{\odot}$ at $z=0.3$ and
sources at $z_
S=1$,
$\theta_E=30$ arcsec.
\end{itemize}
\noindent When the source if slightly off-centered with respect to the
axis defines by
the lens and the observed, a circular lens produces two elongated arcs
diametrally
opposite with respect to the center of the lens.
\begin{figure}
\caption{\label{lensconfig.eps} {\it Simple lens configuration which explains the basic
notations used in this lecture (see Bernardeau for more details).
}
}
\end{figure}
%

\subsection{Critical mass }
By definition, the magnification matrix is given by $A=d \thetag_S /d
\thetag_I$. This defines the convergence, $\kappa$, and the
shear,$(\gamma_1,\gamma_2)$,
\begin{equation}
\left(
\begin{array}{cc}
1-\kappa-\gamma_1 & -\gamma_2 \\
-\gamma_2 & 1-\kappa+\gamma_1
\end{array}
\right)
=\left(
\begin{array}{cc}
1-\partial_{xx}\varphi & -\partial_{xy}\varphi \\
-\partial_{xy}\varphi & 1-\partial_{yy}
\end{array}
\right) \ ,
\end{equation}
\noindent where ``$\partial_{ij}$'' denotes the partial derivative of
the
projected potential with respect to the coordinates $ij$.  The {\sl
magnification}
is given by
\begin{equation}
\mu={1 \over \left|{\rm det}A\right|}={1 \over
\left|\left(1-\kappa^2\right)-\gamma^2\right|} \ ,
\end{equation}
where the shear amplitude
 $\gamma=\sqrt{\gamma_1^2+\gamma_2^2}$  expresses the
 amplitude of the anisotropic magnification.  On the other hand,
 $\kappa$ expresses the isotropic part of the magnification.  In is
worth noting
 that  $2\kappa=\Delta \varphi=\Sigma /\Sigma_{crit}$, is directly
related to the projected mass density. It may happen that the
determinant vanishes. It
that case, the magnification becomes infinite. From an observational
point of view,
 these cases correspond to the formation of very extended images, like
Einstein rings
or giant arcs. The points of the image plane
  where the magnification is infinite are called
 {\sl critical lines}. To these critical lines correspond points in the
source plane
which
are called  {\sl caustic lines}. \\
The critical density is
\begin{equation}
\Sigma_{crit}={c^2 \over 4 \pi G}{D_{OS} \over D_{LS} D_{OL}}
\end{equation}
expresses the capability of a gravitational  system to produce strong
lensing
effect ($\Sigma_{crit} \geq 1$). If we scale  to $2 H_0 /c^2$ in order
to define the ``normalized'' angular distances $d_{ij}$, then
\begin{equation}
\Sigma_{crit} \approx 0.1 \left({H_0 \over 50 {\rm
km/sec/Mpc}}\right){d_{os} \over
d_{ls} d_{ol}} \ {\rm g.cm}^{-2}
\end{equation}
For example, for a lensing-cluster at redshift $z_L=0.3$ and lensed
sources at redshift
$z_S=1$, $d_{os}/(d_{ls}d_{ol})\approx 3$.
If the cluster can be modeled by a
isothermal sphere with a core radius $R_c$ and with $M(R_c)=2 \times
10^{14} \ M_{\odot}$,
then
\begin{itemize}
\item For $R_c$=250 kpc, $\Sigma_{crit}$=0.05  g.cm$^{-2}$,
\item For $R_c$=50  kpc, $\Sigma_{crit}$=1.  g.cm$^{-2}$ \ .
\end{itemize}
Hence, the existence of giant arcs in clusters implies that clusters
should be
strongly concentrated.

\subsection{Relation with observable quantities}
Since, to
 first approximation, faint galaxies look like ellipses, their
shapes can be expressed as function of their weighted
second moments,
\begin{equation}
M_{ij}=\displaystyle{ {\int S(\thetag)
(\theta_i-\theta^C_i)(\theta_j-\theta^C_j
)
d^2\theta \over \int S(\thetag) d^2\theta }} \ ,
\end{equation}
where the subscripts $i$ $j$ denote the
   coordinates $\thetag$ in the
 source and the image planes, $S(\thetag)$ is the surface brightness of
the source and
$\thetag^C$
is the center of the source. \\
Since
gravitational lensing effect does not change the surface brightness
 of the source (Etherington 1933),
 then, if one assumes that
 the magnification matrix is constant across the image,
  the relation between
the shape of the source, $M^S$ and the lensed image, $M^I$ is
\begin{equation}
M^I=A^{-1} \ M^S \  A^{-1}
\end{equation}
Therefore, the gravitational lensing effect
 transforms a circular source
 into an ellipse.
 Its axis ratio is given by the ratio of the two
eigenvalues of the magnification matrix. The shape of the lensed
galaxies
can then provide information about these quantities.
 However, though $\gamma_1$ and $\gamma_2$ describe the anisotropic
distortion
 of the magnification, they are not directly related to
observables  (except in the weak shear regime).
 The {\it reduced complex shear}, $\boldg$, and the complex
polarization (or distortion), $\deltag$,
\begin{equation}
\boldg={\gammag \over (1-\kappa)} \ \ \ ; \ \ \
\deltag= { 2 g \over 1 + \vert \boldg\vert^2} ={2 \gammag (1-\kappa)
\over
(1-\kappa)^2+\vert \gammag \vert^2} \ ,
\end{equation}
\noindent are more relevant quantities 
because $\deltag$ can be expressed in terms of the observed major and
minor axes $a^I$ and $b^I$
of the image, $I$, of a circular source $S$:
\begin{equation}
{ a^2 - b^2 \over a^2 +b^2} = \vert \deltag \vert
\end{equation}
In this case, the 2 components of the complex polarization
 write:
\begin{equation}
\delta_1={M_{11}-M_{22} \over Tr(M)} \ \ \ ; \ \ \ \delta_2={ 2 M_{12}
\over Tr(
M)} \ ,
\end{equation}
 where $Tr(M)$ is the trace of the magnification matrix.
For non-circular sources,
 it is possible to relate the ellipticity of
the image $\epsilong^I$ to the ellipticity of the lensed source,
$\epsilong^S$. In the
general case, it depends on the sign of $Det(A)$ (that is the position
 of the source with respect to the caustic lines)
 which
expresses whether images are radially or tangentially elongated:
\begin{equation}
\epsilong^I={1 + b^I / a^I \over 1 + b^I / a^I} e^{2 i \vartheta} =
{\epsilong^S + \boldg \over 1 - \boldg^* \epsilong^S} \ {\rm for}
 \ Det(A)>0 \ ,
\end{equation}
and
\begin{equation}
\epsilong^I= {1+ \epsilong^{S*} \boldg \over \epsilong^{S*} + \boldg^*}
\ {\rm for}  \ Det(A)<0 \ .
\end{equation}

\section{Academic examples}

\subsection{The singular isothermal sphere}

The projected potential at radius $r$, $\varphi(r)$, of a singular
isothermal sphere with
 3-dimension velocity dispersion $\sigma$  is
\begin{equation}
\varphi=4 \pi {\sigma^2 \over c^2}{D_{LS} \over D_{OS}} r
\end{equation}
and the images are described by the lensing equation
\begin{equation}
\thetag_S=\thetag_I- 4 \pi {\sigma^2 \over c^2}{D_{LS} \over D_{OS}}
{\thetag_I
\over
\left|\thetag_I\right|} \
\end{equation}
\noindent and the magnification matrix
\begin{equation}
\left(
\begin{array}{cc}
1 & 0 \\
0 & \displaystyle{1- 4 \pi  {\sigma^2 \over c^2}{D_{LS} \over D_{OS}} {1
\over
\left|\thetag_I\right|} }
\end{array}
\right)
\end{equation}
\noindent Hence, there is only one critical line which is given by the
Einstein
radius
\begin{equation}
\theta_{SIS}=4 \pi  {\sigma^2 \over c^2}
{D_{LS} \over D_{OS}} \approx 16" \left({\sigma
 \over 1000 {\rm km.sec}^{-1}} \right)^2
\end{equation}
\noindent for an Einstein de Sitter universe, with  $z_L=0.3$ and
$z_S=1$.  The total mass included within the radius $\theta$ is then
\begin{equation}
M\left(\theta\right)=
5.7 \times 10^{13} \ M_{\odot} h_{50}^{-1} \left({\theta \over 16"}\right)
\left({\sigma  \over 1000 {\rm
km.sec}^{-1}}\right)^2
\end{equation}
\noindent From Eq.(28) and (29), it is obvious that the magnification
writes
\begin{equation}
\mu(\theta_I)={\theta_I \over \theta_I-\theta_{SIS}}
\end{equation}
\noindent The singular isothermal sphere permits to keep in mind the
various properties of gravitational lenses 
 and order of magnitude estimates of the associated  physical quantities.  However, more
complex lenses
 can
produce somewhat different configurations of strong lensing. Some
example are shown in
Figure 2, where a series of arcs generated by an elliptical potential
well is
 displayed. We see that various kind of multiple images can be produced,
with some
strange radial arcs for some configurations. Each panel can be found in
the universe
 so this kind of template of lens features can be helpful for the
understanding
 of
 the lens modeling. This has been extensively used, in particular on HST
images
(see
 Fort \& Mellier 1994, Kneib et al 1996, Natarajan \& Kneib 1997,
 Mellier 1998 and references therein).
\begin{figure}
\caption{\label{panelellipsecaustic.eps} {\it Gravitational distortion
 produced  
 by an elliptical potential as a function of source
position. The top left panel shows the shape of the source in the source
plane. The second panel shows 10 positions of the source in its source
plane (referenced from 1 to 10) with respect to a simulated cluster
lens. The thin lines
 show the inner and outer caustics. Panels 3 to 12 show
the inner and outer critical lines and the shape of the image(s) of the
lensed source. Positions 6 and 7  correspond to cusp
configurations, and position 9 is typical fold configuration. On the
fifth panel we see two inner merging images forming a typical radial arc
 (from Kneib 1993, PhD thesis).
}
}
\end{figure}
%

\subsection{The general case of an axially symmetric lens}
Assuming the rescaled angle writes (see Bernardeau, this proceedings)
\begin{equation}
\alphag=\nablag\varphi \ \ \ , \ \ \ \Delta\varphi=2\kappa(x) \ ,
\end{equation}
\noindent for an axially symmetric potential, we have 
\begin{equation}
\alpha={d \varphi \over dx} \ \ \ , \ \ \ {1 \over x}{d \over
dx}\left(x{d \varphi \over dx}\right)=2\kappa(x) \ ,
\end{equation}
\noindent and the mass at radius $x$ is 
 $m\left(\xg\right)=m\left(\left|\xg\right|\right) =m(x)$.
\noindent Hence
\begin{equation}
{d m \over dx}= 2x \kappa(x)  \ \ \Longrightarrow \ \ {d \varphi \over
dx} ={2 \over x} \int_0^x \kappa\left(x'\right) x' dx' \ .
\end{equation}
Therefore, in general the magnification matrix writes
\begin{equation}
A={\cal I}-{m\left(x\right) \over x^4}
\left(
\begin{array}{cc}
x_2^2-x_1^2 & -2 x_1x_2 \\
-2 x_1x_2 & x_1^2-x_2^2 
\end{array}
\right)
-{1 \over x^3} \left({dm(x) \over dx}\right)
\left(
\begin{array}{cc}
x_1^2 & x_x x_2\\
x_x x_2& x_2^2
\end{array}
\right) \ ,
\end{equation}
\noindent where ${\cal I}$ is the identity matrix.  The expression of the
shear terms is then immediate, and the total magnification wirtes:
\begin{equation}
Det\left(A\right)=
\left(1-{m \over
x^2}\right)
\left(1-\displaystyle{ {d\over dx }\left( {m(x)\over x}\right) }
\right)
\end{equation}
\noindent where the first term on the right-hand part provides the
position of the tangential arcs, whereas the second term provides   
the location of the radial arcs. These equations are useful in 
order to compute the lensing  properties of  mass distributions, 
like the universal 
profile discussed in Section 4.1.2 \ .

\section{Astrophysical examples}

\subsection{Measuring the masses of clusters of galaxies}

\subsubsection{The case of MS2137-23}

Strong lensing refers to lensing configurations where the source is
located close to
a caustic line and produces arcs or rings. In addition to the accurate
estimate of the
total mass, strong lensing features can probe some details on the mass
density
profile. \\
A nice example is the case of the lensing cluster MS2137-23
which contains a
radial and a tangential arc.  A simple
investigation already permits to have some constraints on the lens
configuration. First,
since there is no counter-arc associated to the tangential arc on the
diametrally
opposite side of the cluster center, it is unlikely that the projected
mass density
is circularly distributed.  Second, since the tangential arc and the
radial arc are
on the same side, it is unlikely that they are images of the same
source. An interesting
tentative to model the lens 
 is to assume that the dark matter follows the light
distribution of the
central galaxy (see Figure 3).  The isophotes then provide the center, the orientation
and the
ellipticity of light as constraints to  dark matter. Let us assume
that the projected
potential as the following  geometry
\begin{equation}
\varphi(r,\theta)= \varphi_0 \sqrt{1+\left({r \over
r_c}\right)^2\left(1-\epsilon
 {\rm cos}\left(2\theta\right)\right)}
\end{equation}
\noindent where $r_c$ is the core radius, $\epsilon$ the ellipticity,
and $\varphi_0$
the depth of the potential, respectively. For an elliptical lens, the
magnification matrix can be
written
\begin{equation}
\left(
\begin{array}{cc}
1 -\displaystyle{{1\over r}{\partial \varphi \over \partial r}-{1\over
r^2}{\partial^2
\varphi \over \partial \theta^2}} & 0 \\
0 & \displaystyle{1-{\partial^2 \varphi \over \partial r^2} }
\end{array}
\right)
\end{equation}
\noindent which gives immediately the position or the radial and
tangential arcs,
\begin{equation}
\left[1-{\partial^2 \varphi \over \partial r^2}\right]_R=0 \ \ ; \ \
\left[1-{1\over r}{\partial \varphi \over \partial r}-{1\over
r^2}{\partial^2
\varphi \over \partial \theta^2}\right]_T=0
\end{equation}
\noindent From these two constraints, we can have a useful information
of the
core radius itself (Mellier et al 1993):
\begin{equation}
\left({r_R \over r_c}\right)+1 = \left({D_{OL_R} \over D_{OS_R}}\right)
\left({D_{OS_T} \over D_{OL_T}}\right) \left[\left({r_T \over
r_c}\right)^2+1\right]^{3/2}
\end{equation}
\noindent where the subscripts $R$ and $T$ denote the radial and
tangential arc
respectively.  Hence, provided the redshifts of the lens and the two
sources are
 known,
one can infer the core radius, or more generally the steepness of the
dark matter
density. \\
\begin{figure}
\vskip -2cm
\hskip 3truecm
\caption{{\it Model of MS2137-23. This cluster is at redshift 0.33 and shows
a tangential arc and the first radial arc ever detected. Up to now it is
the most constrained cluster and the first one where counter images were
predicted before being observed (Mellier et al 1993). The external and
the dark internal solid lines are the critical lines. The internal grey
ellipse and the diamond are the caustic lines. The thin isocontours
shows the positions of the arcs and their counter images.}
}
\end{figure}
The modeling of the central regions of clusters of galaxies shows that
the geometry of the dark matter follows the geometry of the diffuse
light associated to the dominant galaxies. This is a remarkable success
of the strong lensing analysis. However, there are still some
uncertainties about the total mass which does not agree with the mass
inferred from X-ray analysis. The mass-to-light ratio ranges in the same
domain ($\approx$ 100-300), but one can find frequently a factor of two
difference. 
\begin{table}
\begin{center}
{\small
\caption{Results obtained from strong lensing analyses of
clusters. These are only few examples. $\sigma_{obs}$ is the 
velocity dispersion obtained from spectroscopy of cluster galaxies. 
$\sigma_{DM}$ is the velocity dispersion of the dark matter obtained
from the modeling of the lens.
}
\bigskip
\begin{tabular}{lccccc}\hline\\
\bigskip
Cluster& $z$ & $\sigma_{obs}$ &  $\sigma_{DM}$ & M/L & Scale  \\
 & & (kms$^{-1}$)  & (kms$^{-1}$)& ($h_{100}$) & ($h^{-1}_{100}$ Mpc)
\\
\hline
MS2137-23 & 0.33 & - &900-1200& 680-280 & 0.5 )\\
A2218 & 0.18 &1370&  $\approx$ 1000& 200 & 0.9 \\
A2390 & 0.23 & 1090& 1260 &250 & 0.4 \\
Cl0024+17 & 0.39  & 1250 &1300& $>$200& 0.5 \\
A370 & 0.38 & 1370&850 &$>$150 &0.5 \\
Cl0500-24 & 0.316 & -&$<$1200 &$<$600 &0.5 \\
\\

\hline
\end{tabular}
}
\end{center}
\end{table}

\subsubsection{The X-ray/lensing mass discrepancy}
As it is illustrated in Figure 4, it turns out 
  that the peaks of dark matter revealed by giant arcs in 
  lensing-clusters  correspond to the the peaks if  X-ray emission 
 as well as the location of the brightest components 
 of optical images.
 On the other hand, Miralda-Escud\'e \& Babul (1995) have 
 pointed out an  apparent contradiction
between the mass estimated from X-ray data and the lensing mass
($M_{lensing} \approx 2-3 M_X$).  Further works done by  many groups 
   lead to somewhat inconclusive statements about this contradiction.
  B\"ohringer et al (1998) find an excellent
agreement between X-ray and lensing masses in A2390 which confirms
 the  view claimed by Pierre et al (1996); Gioia et al (1998)
show that the disagreement reaches a factor of 2 at least in
MS0440+0204; Schindler et al (1997) find  a factor of 2-3  discrepancy
for the massive cluster RXJ1347.4-1145, but Sahu et al (1998) claim
that the disagreement is marginal and may not exist; Ota et al (1998)
and
 Wu \& Fang (1997) agree that there are
 important discrepancies in A370, Cl0500-24 and Cl2244-02.\\
 There is still no definitive
interpretation of these contradictory results. It could be that the
 modeling of the gravitational mass from the X-ray distribution is not
as simple.
By comparing
 the geometry of the X-ray isophotes of A2218 to the mass
 isodensity contours of the reconstruction,
Kneib et al (1995) found  significant discrepancies
in the innermost parts. The numerous substructures  visible in the
 X-ray image have
orientations which do not follow the projected mass density.
They interpret these features as shocks produced by the in-falling X-ray
gas,
 which implies that the current
 description of the dynamical stage of the inner X-ray gas is
oversimplified.
Recent ASCA
observations of three lensing-clusters corroborate the view that
substructures are the major source of uncertainties.
\\
\begin{figure}
\caption{\label{panelela370.eps} {\it Analysis of the the matter distribution in the
cluster
A370 using strong lensing features (arcs and arclets). The top
right panel shows the R-light distribution from the isoluminosity
contours of galaxies.  The top left panel shows the number density
contours. The bottom right is a B CCD image of the clusters with
the X-ray luminosity contours overlayed. The arclets directly show 
 the shape of the projected mass density. In
particular, we see that the arclet pattern indicates the presence of an
extension toward the eastern region, which is also seen in the
isoluminosity contours in the R-band and the X-ray maps. This clearly
shows that in the center of this cluster light from galaxies and from
the
hot gas trace the mass. Note the arc reconstruction and the mass
model in the fourth panel (the triplet B2-B3-B4 is discussed
by Kneib et al. 1993).
}
}
\end{figure}
In order to study this possibility in more details,
  Smail et al (1997) and Allen (1998) have performed a detailed
comparison
between the lensing mass and X-ray mass for a significant number of
lensing clusters. Both works conclude that the substructures have a
significant impact on the estimate of
X-ray mass. More remarkably,
the X-ray clusters where cooling flows are present do not show a
significant discrepancy with X-ray mass, whereas the others X-ray
clusters do (Allen 1998).  This confirms that the discrepancy is
certainly due to
 wrong assumptions on the physical state of the gas. The interpretation
of this dichotomy in cluster samples may be the following. Clusters with
cooling flows are compact and rich systems which probably have 
probably virialised and have a well-defined 
 relaxed core. Therefore, when removing
the  cooling flow contribution, the assumptions that the gas is in 
hydrostatic equilibrium is fully satisfied. Conversely, non-cooling flow
 clusters are generally poor,  do have lot of substructures and  
 no very dense core dominates the cluster yet. For these systems, 
 the gas cannot be described simply (simple geometry, hydrostatic 
equilibrium) and the oversimplification of its dynamical stage 
 produces a wrong mass estimator.   This interpretation needs 
 further confirmations. However, from these  two
 studies we now have the feeling that we are now close to understand
the origin of the X-ray and lensing discrepancy.\\
An alternative has been
suggested by  Navarro, Frenk \& White (1997) who proposed that the
analytical models
 currently used for modeling mass distributions  may be inappropriate
 (hereafter NFW).
They argue that the universal profile of the
mass distribution produced in numerical simulations of hierarchical
clustering may reconcile the lensing and X-ray masses. This kind 
of profile must be considered seriously because 
 the universal profile is a natural
outcome
from the simulations which does not use external prescriptions.  \\
For the reader who want to look more deeply at the lensing properties of 
 this profile, it is 
 interesting to describe the NFW properties into more details.  
Let us assume that the cluster 3-dimension mass density has the NFW 
shape:
\begin{equation}
\rho(x)={\rho_s \over x \left(1+x\right)^2} \ \ , {\rm with} \ \ 
x={r \over r_s} \ .
\end{equation}
\noindent The projected mass density of the  NFW profile is (use 
equations of  Sect.  3.2)
\begin{equation}
\Sigma(x)=\int_{-\infty}^{+\infty}\rho(x) dz=2 \rho_s r_s {f(x) \over
x^2-1}
\end{equation}
\noindent where
\begin{equation}
f(x)=
\left\lbrace
\begin{array}{ll}
 1-\displaystyle{{2 \over \sqrt{x^2-1}}}{\rm arctg}\left(\displaystyle{
\sqrt{{x-1 \over x+1}}}\right) & \left(x>1\right) \\
 0 & \left(x=1\right) \\
 1-\displaystyle{{2 \over \sqrt{1-x^2}}}{\rm argth}\left(\displaystyle{
\sqrt{{1-x \over 1+1}}}\right) & \left(x<1\right)
\end{array}
\right.
\end{equation}
\noindent  Then the convergence and the mass write as follows:
\begin{equation}
\kappa(x)={2 \rho_s r_s \over \Sigma_{crit}}{f(x) \over x^2-1} \ \ 
, \ \ m(x)=2 \kappa g(x) \ \ ,
\end{equation}
\noindent where
\begin{equation}
g(x)={\rm ln}\left({x \over 2}\right)+
\left\lbrace
\begin{array}{ll}
 \displaystyle{{2 \over \sqrt{x^2-1}}}{\rm arctg}\left(\displaystyle{
\sqrt{{x-1 \over x+1}}}\right) & \left(x>1\right) \\
 1 & \left(x=1\right) \\
 \displaystyle{{2 \over \sqrt{1-x^2}}}{\rm argth}\left(\displaystyle{
\sqrt{{1-x \over 1+1}}}\right) & \left(x<1\right)
\end{array}
\right. \ .
\end{equation}
\noindent Now, from the analytical shape of $M(x)$, it is clear that 
 $\left(d/dx\right)\left(M/x\right)\longrightarrow \infty $ when 
 $x\longrightarrow 0$ and that $\left(d/dx\right)\left(M/x\right)
\longrightarrow 0$ when $x\longrightarrow \infty$. Therefore, 
 there is always a radial line, whatever $\rho_s$ and $r_s$. \\
In the fortunate cases of $MS2137-23$ and $A370$ which 
both have a tangential and a radial arc, one can infer $r_s$ and $\rho_s$
from the analysis of the tangential and radial arcs. As far as the 
redshifts of the two arcs are known, then the ratio of the critical mass
 density at these two positions is, 
\begin{equation}
{\Sigma_{Crit,r} \over \Sigma_{Crit,t}}=
\left({x_t^2 \over g\left(x_t\right)}\right) \left[{1 \over R_{rt}} 
\left({dg\left(R_{rt}x_t\right)\over dx} \right) -\left({1\over R_{rt}x_t }\right)^2
g\left(R_{rt}x_t\right) \right] \ ,
\end{equation}
\noindent where $R_{rt}=x_r/x_t$.  Once 
 used jointly,  the positions and redshifts of the two arcs and 
 the two independent equations, permit to infer $r_s$ and $\rho_s$. 
\\
For example, for $A370$ (see the radial arc in Figure 8), we have
\begin{itemize}
\item $z_{lens}=0.375$, $z_t=0.724$ (measured), $z_r \approx 1.5$ (assumed),
\item $R_{rt}=0.7$
\item $\Sigma_{Crit,t}=$ 1.4 $h_{100}$ g.cm$^{-2}$, \ \ \ \ $\Sigma_s=\rho_s
r_s=$0.28 $h_{100}$ g.cm$^{-2}$
\item $r_s\approx$ 250 $h_{100}^{-1}$ kpc,
\item and the overdensity $\delta_c=\rho(0)/\rho_{critic}=2\times 10^4$, if
$\Omega_0=1$.
\end{itemize}
It is worth noting that the statement  that NFW profiles do predict  radial arcs 
 contradicts the general view that their existence  
  rules out mass profiles with
singularity.  \\
 Despite this interesting prediction which makes the NFW profile 
 even more attractive, Bartelmann (1996) has shown that the caustics 
produced by
this profile predict that radial arcs should be
thicker than observed in MS2137-23
 and in A370, unless the sources
 are very thin ($\approx 0.6$ arcsecond for MS2137-23).
 This is not   a strong argument against the universal profile
  because this is possible
in view of the shapes of some faint galaxies observed with HST that some
distant galaxies are indeed very thin. But it is surprising that no radial
arcs produced by ``thick galaxies''
have  been detected so far. Even a selection bias would probably
favor the observation of large sources rather than small thin and hardly
visible ones.  
 
\subsubsection{Clusters from weak lensing analysis}
Deep images of lensing-clusters
  show many weakly lensed galaxies having a correlated
 distribution of ellipticity/orientation which maps the projected mass
density (Fort et al 1988 and Figure 5). The  first attempt to use this distribution
of arclets as a probe of dark matter has been done by
 Tyson et al (1990), but the  rigorous inversion technique
was first proposed by  Kaiser \& Squires (1993). \\
The weak lensing 
 analysis starts from the following hypotheses:
\begin{itemize}
\item Assume that the orientation of the sources is isotropic.
\item Assume that the orientation of the source is not correlated to
their ellipticity.
\item Assume that the redshift distribution of sources is known.
\end{itemize}
Then it proceeds as follows:
\begin{itemize}
\item Measure the averaged ellipticity and orientation of the galaxies 
 inside all subareas of the field. 
\item Produce a  (ellipticity, orientation) map (see Figure 6).
\item Provide a relation between the (ellipticity, orientation) and 
the components of the shear.
\item Provide a relation between the shear and the mass density.
\item Provide a relation between the shape of the source and the shape
of the image.
\end{itemize}
\noindent The hypotheses can be rather well controled (in principle) 
from the observations of unlensed areas. Ellipticities of field galaxies 
 provide control fields for the first and second assumptions. Spectroscopic 
surveys and photometric redshifts allows to model the redshift distribution 
of galaxies. The procedure itself requires technical analysis of the data
 (see Mellier 1998 for the technical issues) and 
 theoretical relations provided by gravitational lensing theory.  The crucial 
points are the relations between shear, mass density and geometry of the lensed galaxies. 
 This is done by combining the following equations:
\begin{equation}
\left\lbrace
\begin{array}{l}
\gamma=\gamma_1+i\gamma_2=\displaystyle{{1 \over
2}\left(\partial_{xx}-\partial_{yy}\right)\varphi+i\partial_{xy}\varphi}
\\
\kappa=\displaystyle{{1\over 2}\left(\partial_{xx}+
\partial_{yy}\right)\varphi}\\
\varphi=\displaystyle{{1 \over \pi}\int\kappa\left(\thetag'\right) {\rm
ln}\left(\left|\thetag-\thetag'\right|\right)d\thetag'} \ ,
\end{array}
\right.
\end{equation}
from which one can express the complex shear as
a function of the convergence, $\kappa$ (see
 Seitz \& Schneider 1996 and references
therein):
\begin{equation}
\gammag(\thetag) = {1 \over \pi} \int
{\cal D}(\thetag-\thetag')
\ \kappa(\thetag') d^2\theta'  \ ,
\end{equation}
where
\begin{equation}
{\cal D}(\thetag-\thetag')={(\theta_2-\theta_2')^2 -
(\theta_1-\theta_1')^2- 2i
(\theta_1-\theta_1')(\theta_2-\theta_2') \over \vert (\thetag-\thetag')
\vert^4
} \ .
\end{equation}
This equation can be inverted in order to express the
projected mass density, or equivalently $\kappa$, as function of the
shear:
\begin{equation}
\kappa(\thetag)= {1 \over \pi}  \int \Re[{\cal D}^*(\thetag-\thetag')
\gammag(\thetag')]  \ d^2\theta' \ + \kappa_0  \ ,
\end{equation}
where $\Re$ denotes the real part.
 Finally, from Eq.(9-12) we can express the shear as a function
of the complex
ellipticity. Hence, if the background ellipticity distribution is
randomly
distributed, then $<\vert \epsilong^S \vert>=0$ and
\begin{figure}
\caption{\label{unlensedlensed.eps} {\it  Distortion field generated by a lens. The top
panel shows
 the grid of randomly distributed
background sources as it would be seen in the absence of the
lens. The projected number density corresponds to very deep exposure, 
similar to the HDF. The bottom panel shows the
same population once they are distorted by a foreground
(invisible) circular cluster
with a typical velocity dispersion of $1300 \ kms^{-1}$.
The geometrical signature of
 the cluster is clearly visible. The potential can be
recovered by using the formalism defined in part 4. In this
simulation, the sources are at $z=1.3$, and the cluster at $z=0.15$.
}
}
\end{figure}
\begin{equation}
<\vert \epsilong^I \vert>=  \vert \boldg \vert = {\vert \gammag \vert
 \over 1 - \kappa}
\end{equation}
 (Schramm \& Kayser 1995). In the most extreme case, when $\kappa<<1$
(the linear regime),
 $<\vert \epsilong^I \vert> \approx  \vert \gammag \vert$, and
 therefore, the projected mass density can be recovered directly from
the
measurement of the ellipticities of the lensed galaxies.  \\
Alternatively, one can measure the total mass within a circular radius
using the {\it Aperture densitometry} technique (or the ``$\zeta$-{\it
statistics}''),
which
consists in computing the difference between
 the mean projected mass densities within a radius $r_1$ and
  within an annulus $(r_2-r_1)$ (Fahlman et al 1994, Kaiser 1995)
as
function of the {\it tangential shear}, $\gamma_t=\gamma_1 {\rm
cos}(2\vartheta)+\gamma_2 {\rm sin}(2\vartheta)$, averaged inside
  the ring. Let us denote $\bar{\kappa}$ the averaged value of $\kappa$
inside the loop of a circle with radius $r$ and $\left<\kappa\right>_{\theta}$ the
averaged value of $\kappa$ over the loop. We have
\begin{equation}
\bar{\kappa}={1 \over \pi r^2}\int_0^{2\pi}\int_0^r
\kappa\left(r',\theta'\right) r'dr'd\theta'
\end{equation}
\noindent and
\begin{equation}
{d \bar{\kappa} \over dr}=-{2 \over r}{1 \over \pi r^2}
\int_0^{2\pi}\int_0^r \kappa\left(r',\theta'\right) r'dr'd\theta'
+{1 \over \pi r^2} {d \over dr}\left(\int_0^{2\pi}\int_0^r
\kappa\left(r',\theta'\right) r'dr'd\theta'\right)
\end{equation}
\noindent therefore,
\begin{equation}
{d \bar{\kappa} \over dr}=-{2 \over r}\bar{\kappa}+{2 \over
r}\left<\kappa\right>_{\theta} \ .
\end{equation}
\noindent or equivalently
\begin{equation}
{1 \over 2} {d\kappa \over d{\rm
ln}r}=\left<\kappa\right>_{\theta}-\bar{\kappa}
\end{equation}
\noindent Now, the mean tangential shear writes
\begin{equation}
\left<\gamma_t\right>={1 \over 2 \pi}\int_0^{2\pi} \gamma_td\theta'
\end{equation}
\noindent whereas in polar coordinates
\begin{equation}
\gamma_t={1 \over 2}\left(\partial_{rr}\varphi-\partial_{ll}\varphi\right)=
\partial_{rr}-\kappa \ .
\end{equation}
\noindent Therefore
\begin{equation}
\left<\gamma_t\right>={d \over dr}\left(\int_0^{2\pi}\partial_r\varphi
{d\theta' \over 2 \pi}\right)-\left<\kappa\right>_{\theta} \ ,
\end{equation}
\noindent which implies that
\begin{equation}
\left<\gamma_t\right>=-{1 \over r^2}\left<r{d \varphi \over
dr}\right>_{\theta}+{1 \over r}{d \over dr}\left(\left<r
{d\varphi \over dr}\right>_{\theta}\right)-\left<\kappa\right>_{\theta} \ .
\end{equation}
\noindent Therefore, from Eq.(42), 
 $\left<\gamma_t\right>$ is related to $\bar{\kappa}$ by this
simple relation:
\begin{equation}
\left<\gamma_t\right>={1 \over 2} {d\bar{\kappa} \over d{\rm ln}r}
\end{equation}
\noindent which provides the $\zeta$ estimator:
\begin{equation}
\zeta(r_1,r_2)=<\kappa(r_1)>-<\kappa(r_1,r_2)>={2 \over 1-r_1^2/r_2^2}
\
\int_{r_1}^{r_2} <\gamma_t>
 d {\rm ln}r \ .
\end{equation}
This  quite robust mass estimator minimizes
 the contamination by foreground and cluster galaxies and permits a
simple check that the
signal is produced by shear,  simply by changing $\gamma_1$ in
$\gamma_2$ and
$\gamma_2$
in $-\gamma_1$ which should cancel out the true shear signal.\\
The generalization to the non-linear regime (Kaiser 1995 and
 Seitz \& Schneider 1996) can be done
 by solving the integral equation obtained from Eq.(38) where
 $\gammag$ is replaced by $(1-\kappa)\boldg$. Alternatively one can
 use the fact that both $\kappa$ and $\gammag$
 depend on second derivatives of the projected gravitational potential
$\varphi$:
\begin{equation}
\left
\lbrace
\begin{array}{l}
\displaystyle{{\partial \kappa \over \partial x_1}={\partial \gamma_1 \over \partial
x_1}+ {\partial \gamma_2 \over \partial x_2}}\\
 \\
\displaystyle{{\partial \kappa \over \partial x_2}={\partial \gamma_2 \over \partial
x_1}-{\partial \gamma_1 \over \partial x_2}}
\end{array}
\right.
\end{equation}
\noindent  which permits to recover the mass density by
 this relation:
\begin{equation}
\nabla {\rm log}(1-\kappa)= {1 \over 1 -\vert \boldg \vert^2}
 \ .
\left(
\begin{array}{cc}
1+g_1 & -g_2 \\
-g_2   & 1-g_1 \\
\end{array}\right) \  \
\left(
\begin{array}{c}
\partial_1 g_1 + \partial_2 g_2 \\
\partial_1 g_2 - \partial_2 g_1 \\
\end{array}
\right)
\end{equation}
Both Eq.(36) and Eq.(51) express the same relation between $\kappa$ and
$\gammag$ and can be used to reconstruct the projected mass
density.  \\
Although mass reconstruction is now as a robust technique
(see the comparison of various algorithms in Mellier 1998), the
mass distribution recovered is not unique because the addition of a lens
plane with constant mass density keeps the distortion of the
galaxies unchanged. Furthermore, the inversion only uses the ellipticity
of the
galaxies regardless of their dimension, so that changing  $(1-\kappa)$
in
$\lambda(1-\kappa)$ and $\gammag$ in $\lambda$ $\gammag$ keeps $\boldg$
invariant.
This is the so-called {\it mass sheet degeneracy}
 (Gorenstein et al 1988).
  \\
\begin{figure}
\caption{\label{cl0024shear.eps} {\it Detection of the shear field around Cl0024+1654.
The
figure is composed of two deep CCD images obtained at CFHT. 
 The small field on the right is the central  region of the cluster. 
 The off-centered field on the
left covers a much larger field and has been observed in order to detect 
 the mass distribution at the periphery of the cluster. The thick full lines indicate the
local average ellipticity. Each line displays the amplitude and the orientation 
 of the distortion.  The pattern is
typical of a coherent gravitational  shear produced by the
 mass of the cluster. Note also the perturbation of the shear field in the
upper left. This effect is due to a secondary deflector which locally modifies
 the shear field.
}
}
\end{figure}
The degeneracy could in principle be broken
 if the magnification can be measured independently, since
 it is not invariant under the linear transformation mentioned above,
but instead it is reduced but a factor $1/\lambda^2$.
 The magnification can be measured directly
 by  using the
 magnification bias (Broadhurst et al 1995),
  which changes the galaxy number-counts. \\
The magnification bias expresses
the  effects of  the gravitational magnification, which
 increases the flux received from  lensed galaxies and 
  magnifies by the same amount the area of the projected
lensed sky and thus decreases the apparent galaxy number density.
The total amplitude of the magnification bias
depends  on the slope of
the galaxy counts as  a function of magnitude and
on the magnification factor of the lens. For a circular lens, the
 radial galaxy number density of background galaxies writes:
\begin{equation}
   N(<m,r) = N_0(<m) \ \mu(r)^{2.5\alpha-1} \ \approx N_0 \ (1+2
\kappa)^{2.5\alpha-1} \ \ \ \  {\rm if } \ \kappa\ {\rm and} \ \vert
\gammag\vert \ll 1\  ,
\end{equation}
 where  $\mu(r)$ is the magnification, $N_0(<m)$ the
intrinsic (unlensed) number density, obtained from
 galaxy counts in a nearby empty field, and
  $\alpha$  is the intrinsic count slope:
\begin{equation}
\alpha = {{\rm d}logN(<m) \over {\rm d}m }\ .
\end{equation}
A radial magnification bias $N(<m,r)$ shows up only when
the slope $\alpha \not=0.4$; otherwise, the increasing number of
magnified sources is exactly compensated by the apparent field dilatation.
For slopes larger than $0.4$ the magnification bias increases the galaxy
number density, whereas for slopes smaller than $0.4$ the radial density
 will show a depletion.
Hence, no change in the galaxy number density can
be observed for $B(<26)$ galaxies, since
the slope is almost this critical  value (Tyson 1988).
 But it can be detected in the $B>26$, $R>24$ or $I>24$ bands when the slopes
are close to 0.3 (Smail et al 1995).  
The change of the galaxy number density can be used as a direct
measurement of the magnification and can be included in the maximum likelihood
inversion as a direct observable in order to break the
 mass sheet degeneracy .
\\
The    {\it lens parallax method}
 (Bartelmann \& Narayan 1995) which compares
 the angular sizes of lensed galaxies with an unlensed sample can be also 
used as an alternative to beak the degeneracy.
  Another approach consists in using
wide field cameras with a field of view much larger than clusters
of galaxies. In that case  $\kappa$ should  vanish at the boundaries of
the field, so
that the
degeneracy could in principle be broken.\\
\begin{table}
\begin{center}
{\small
\caption{Results obtained from weak lensing analyses of
clusters. The
scale is the typical radial distance with respect to the cluster center.
The last cluster has two values for the M/L ratio. This corresponds to
two extreme redshifts assumed for the lensed population, either $z=3$ or
$z=1.5$. For this case, the two values given for the velocity dispersion
are those inferred when $z=3$ or $z=1.5$ are used.
}
\bigskip
\begin{tabular}{lccccc}\hline\\
\bigskip
Cluster& $z$ & $\sigma_{obs}$ &  $\sigma_{wl}$ & M/L & Scale  \\
 & & (kms$^{-1}$)  & (kms$^{-1}$)& ($h_{100}$) & ($h^{-1}_{100}$ Mpc)
\\
\hline
A2218 & 0.17 & 1370 &-& 310 & 0.1 )\\
A1689 & 0.18 &2400&  1200-1500 & - & 0.5 \\
      &  & &-& 400& 1.0 \\
A2163 & 0.20 & 1680&740-1000  & 300&0.5 \\
A2390 & 0.23 & 1090& $\approx$1000 &320 & 0.5 \\
Cl1455+22 & 0.26  &$\approx$ 700 & - &1080&0.4 \\
AC118 & 0.31 & 1950 & -&370 & 0.15 \\
Cl1358+62 & 0.33  & 910&780 &180 & 0.75\\
MS1224+20 & 0.33  & 770& -& $\approx$ 800& 1.0 \\
Q0957+56  & 0.36  & 715&- & -& 0.5 \\
Cl0024+17 & 0.39  & 1250 &-& 150& 0.15 \\
          & & & 1300& $\approx$900& 1.5\\
Cl0939+47 & 0.41  & 1080 &-&120 & 0.2 \\
          &    &  &-&$\approx$250 & 0.2 \\
Cl0302+17 & 0.42  & 1080& & 80& 0.2 \\
RXJ1347-11 & 0.45  & - &1500& 400& 1.0 \\
3C295 & 0.46 & 1670 &1100-1500 &- &0.5 \\
 &  & &-& 330&0.2 \\
Cl0412-65 & 0.51  &-&-&70 & 0.2 \\
Cl1601+43 & 0.54  &1170 &-&190 & 0.2 \\
Cl0016+16 & 0.55  &1700 &-&180 & 0.2\\
          &       & &740 &740&0.6 \\
Cl0054-27 & 0.56  &-&-&400 & 0.2\\
MS1137+60 & 0.78  &859&-&270 &0.5\\
RXJ1716+67 & 0.81  &1522&-&190 &0.5 \\
MS1054-03 & 0.83 & 1360&1100-2200 &350-1600 &0.5 \\
\\

\hline
\end{tabular}
}
\end{center}
\end{table}
Since 1990, many clusters have been investigated using the weak lensing
inversion, either using ground-based or HST data. They are summarized in
Table 2,
 but the comparison of these results is not straightforward because of
the different observing conditions which produced  each set of data
 and the different mass reconstruction algorithms used by each author.
 Nevertheless, all these studies
show that on scales of about 1 Mpc, the geometry of
mass distributions, the X-ray
distribution and the galaxy distribution are similar
(see Figure 4),
though the ratio
of each component with respect to the others may vary with radius. The
 inferred median M/L
value is about 300, with a trend to increase with radius.
   Contrary
to the strong lensing cases, there is no evidence of discrepancies
between the X-ray mass and the weak lensing mass. It is worth
 noting that the strong lensing mass and the weak lensing mass estimates
are consistent in the region where the amplitude of two regimes
 are very close. This is an indication
that the description of the X-ray gas, and its coupling
 with the dark matter on the
scales corresponding to strong lensing studies is oversimplified,
whereas on larger scales, described by weak lensing analysis, the
detailed description of the gas has no strong impact. \\
The large range of M/L could partly be a result of one of the
issues of the mass reconstruction from weak lensing.  
 In particular, the deviation angle depends on the ratios of the three
angular-diameter distances (see Bernardeau, these proceedings), 
 which depends on  the redshift
 we assume for the sources. For low-redshift lenses, the dependence with
redshift
of the background galaxies
is not important, so the calibration of the mass can be provided with
a reasonable confidence level.  However, distant clusters
 are highly sensitive to the redshift of the sources,
  and it becomes very difficult to scale the total
mass without this information.
\vskip 5truemm
From the investigation of
about 20 clusters, it turns out that the median M/L is lower than 400.
This implies that weak  lensing analyses predict
   $\Omega<0.3$ with a high significance level.
 These constraints
 on $\Omega$ are in good agreement with other observations.
 \\
Another strong statement results from the mass reconstruction obtained
by Luppino \& Kaiser (1997) and Clowe et al (1998) or from the detection
 of giant arcs in very distant clusters (Deltorn et al 1997): massive
clusters
 do exist at redshift $\approx$1 ! This is a strong but 
 reliable statement, though the total mass and the M/L
cannot be
 given with a high accuracy. Therefore,  unless
 unknown important systematics have been disregarded, 
 we now have the first direct observational  evidences
 that high mass-density peaks have generated massive clusters of
galaxies at redshift 1. These promising results
  are corroborated by weak lensing studies around radio sources
and quasars (Mellier 1998).

\subsection{Measuring the masses of galaxies}
Gravitational lensing can also   
provide valuable insight on the halos of galaxies. Since it  works on
all scales, in principle the halos of galactic dark matter
 could be probed from their gravitational lensing effects on
 background galaxies. 

\subsubsection{Einstein rings}
Rings occur when the alignment of the observer, the lens and
the source is almost perfect, and if the source is covering the whole
internal caustic, forming the so called "Einstein ring".
 The first rings were observed around galaxies in radio surveys
(see Refsdal \& Surdej 1994 for a recent review).
 They have provided unique targets to measure
 the mass-to-light ratios and to probe the mass profiles
of  galaxies (Kochanek 1991). In the case of rings, the mass
 of the lensing galaxies can be very well constrained (see for instance
Kochanek 1995 and Table 3),
so  the properties of the
halos inferred from modeling are reliable. The results are somewhat 
reasonable, with typical velocity dispersion and mass-to-light ratio 
in good agreement with other techniques.  However, I would like to stress again
that, due the simplicity of these lens 
configurations and the very strong constraints provided by the 
 size of the ring, these results are very robust.
\begin{table}
\begin{center}
{\small
\caption{Results on Einstein ring analyses. This is not a complete 
survey of the rings detected. I only report on those for which enough
data have been obtained and a model has been presented.
$^{(1)}$ Kochanek 1995, \ $^{(2)}$ Impey et al 1998, \ $^{(3)}$ Warren et al 1998.
}
\bigskip
\begin{tabular}{lcccc}\hline\\
\bigskip
Lens& $z_{lens}$ & $z_{source}$ & $\sigma_{DM}$ &  M/L \\
 & & & (kms$^{-1}$)  & ($h_{100}$) \\
\hline
MG 1654+134$^{(1)}$ & 0.25 & 1.74 & $\approx $ 220 & $\approx $ 20.4 (in B)\\
PG 1115+080$^{(2)}$ & 0.310 & 1.722&  $\approx $ 240 & $\approx $ 8.2 (in I) \\
\ \ \ \ 0047-2808$^{(3)}$  &0.485  &3.595 &$\approx $ 270& - \\
\hline
\end{tabular}
}
\end{center}
\end{table}

\subsubsection{Perturbations near giant arcs}
Perturbations of caustics by intervening masses can locally change
the length and shape
of arcs or locally increase the intensity of unresolved
arc substructures. Dramatic
perturbations could even be responsible for the complete vanishing of
an arc segment. The perturbation of caustics by a smaller interloping
lens can be understood by considering that the magnification matrix
degenerates to a single eigenvector tangent to the critical curve.
Indeed
the distortion of the images of objects close to the critical
line corresponds mainly to a stretching
 along the direction of merging. Therefore,
the angular coordinates of the source can be developed
 in polynomial form along
the direction of merging (Kassiola, Kovner \& Fort 1992).
Figure 7 gives a short description of the effect of
perturbations on two merging images of an extended object
near a fold.  In this case the functional form of the
unperturbed fold is
approximated by a second-order polynom. If a large
  perturbation from a nearby galaxy is added, the
image can be split into many components (Figures 7 and 8)
\\
\begin{figure}
\caption{\label{perturbcaustic.eps} {\it  Gravitational distortion induced by a
perturbation close
to
a giant fold arc. The top left panel shows the formation of two
elongated
images by a fold catastrophe. The vertical segment (A,B) is the length
of the source in the source plane. The images are given by the antecedent
of the parabola (fold caustic) and therefore two images are formed. In
the next  panel we introduce a perturbation represented as the dashed
line which co-adds to the parabola. This perturbation roughly represents
the deviation angle expected from an isothermal sphere with soft core,
assuming its influence is zero beyond a given radius. The difference
between the three configurations is the intensity of the perturbation.
When the intensity is large enough (top right panel)
it can break the nearest image
 and form multiple small images. When the intensity
decreases (bottom left)  the perturbation can break the images but can
also form sub-ellipses of merging sub-images. 
}
}
\end{figure}
%
\begin{figure}
\caption{\label{panelperturb.eps} {\it  Perturbations close 
to giant arcs produced by galaxies.  The top panel is the giant arc 
 in A370. The galaxies with a number are those reported in Table IV. On this
HST images the effects of these galaxies is clear, in particular for the galaxy 
$\#22$. The middle panel shows similar pertubation in Cl0024+17. In particular, 
 the effect of $\#158$ is important. Though the central arc should be twice as long as
the others (prediction of cups arcs), it is clearly smaller. The contraction is
produced by the two galaxies which are located at the top and the bottom of the central
arc. The bottom panel shows, the best lens model (left) of MS2137-23 and 
  an example of the perturbation of the galaxy $\#7$. When two much mass is put
in this galaxy one can see that the giant arc is broken in three sub-arcs. This 
 effect permits to put upper limits on the mass of this galaxy.
}
}
\end{figure}
Large perturbations of
caustics have been used to constrain the galaxies 
 located close to the giants arcs in A370, Cl0024+17, Cl2244 or 
 MS2137-23.  In general, the absence of breaks along a well-defined arc 
 provides robust upper limits to the mass of perturbing galaxies.  
In summary, the results, as those shown in Table 4 are 
 not suprising.  The masses found for these cluster galaxies
 range between $10^{10}$ M$_{\odot}$ and 2 $\times 10^{11}$ M$_{\odot}$, with
typical mass-to-light ratios between 5 and 30.  
\begin{table}
\begin{center}
{\small
\caption{Results cluster galaxies near arcs (perturbations). The 
table gives the constrains on the masses of the halos of galaxies around 
giant arcs which are shown in Figure 8.
}
\bigskip
\begin{tabular}{lcccc}\hline\\
\bigskip
Lens& $z_{lens}$ & $\#$ Galaxy & $\sigma_{DM}$ &  M/L \\
 & & & (kms$^{-1}$)  & ($h_{100}$) \\
\hline
A370 & 0.38 & 201 & -- &  $2.4<M/L<8$ (in B)\\
A370 & 0.38 & 22&  -- &  $4.4<M/L<12.6$ (in B) \\
A370 &0.38  &63 &-- &  $9<M/L<30$ (in B) \\
MS2137-23 &0.33  &7 &$\le$ 190 & -- \\
Cl0024+17 &0.39  &m1 &$\approx$ 180 & -- \\
Cl0024+17 &0.39  &186 &$\approx$ 250 & -- \\
Cl0024+17 &0.39  &m2 &$\approx$ 120 & -- \\
\hline
\end{tabular}
}
\end{center}
\end{table}

\subsubsection{Galaxy-galaxy lensing on field galaxies}
A more promising approach consists in a statistical study of the
  deformation of distant galaxies by foreground galactic halos.
 The galaxy-galaxy lensing analysis uses the correlation
 between the position 
 of foreground galaxies and the orientation 
of background population.
If the correlation is produced by
 the gravitational shear of the foreground halos, then it is
possible to probe their mass, if the redshift distributions of the
foregrounds and the backgrounds are known. \\
Let us define the shape of a galaxy by the vector $\epsilong$ as defined
in Eq.(12).
In the weak lensing regime, the shear simply
  translates in the $\epsilong$ plane
 at the new position  $\epsilong=\epsilong_0+\deltag$, where
$\epsilong_0$ is the intrinsic shape and $\deltag$ the distortion
produced
by the shear. If we assume that the translation is done in the
$x$-direction, then the shape distribution is simply translated:
\begin{equation}
f_{\epsilon}\left(\epsilon_x,\epsilon_y\right)=f^0_{\epsilon}
\left(\epsilon_0,\epsilon_y\right)
=f^0_{\epsilon}\left(\epsilon_x-\delta,\epsilon_y\right) \ ,
\end{equation}
where $f^0_{\epsilon}$ is the intrinsic shape distribution. In that
case,
the orientations of the galaxies with respect to the $x$-axis, $\phi$,
  is modified and the final distribution, averaged over the ellipticity
of the galaxies, is
\begin{equation}
P_{\phi}\left(\phi\right)=\int f^0_{\epsilon} \ \epsilong \ d\epsilong +
\delta {\rm cos}\left(2\phi\right) \int \epsilong \ {d  f^0_{\epsilon}
\over
d \epsilon} d\epsilong
\end{equation}
\noindent that is
\begin{equation}
P_{\phi}\left(\phi\right)={2 \over \pi}\left(1-\left<\delta\right>{\rm
cos}\left(2\phi\right)\left<\epsilong^{-1}\right>\right)
\end{equation}
\noindent Therefore, we expect a deficit of radially-oriented galaxies
and conversely an excess of tangentially-oriented images.  \\
A statistical analysis is
then possible, if one assumes that all the foreground galaxies have
similar halos,
 which can be scaled from observations.
  The procedure is the following.
Assume an analytical shape for the projected potential $\varphi$ 
 (or for the projected mass density $\Sigma$, or the 3-dimension mass
density $\rho$). Then,  
assuming the potential of halos are circular, the deflection angle is 
\begin{equation}
\alpha(r)={2\over c^2} {D_{LS} \over D_{OS}}
\displaystyle{{d\varphi \over dr}}
\end{equation}
\noindent and the polarization writes
\begin{equation}
p(r)=D_{OL} \ r {d \over dr}\left({\alpha(r) \over r}\right)
\end{equation}
\noindent which depends on the typical scale, say $r_c$ and the depth 
of the
potential well ({\it i.e.} for an ellipical galaxy or a cluster of
galaxies  the velocity dispersion of the lens, $\sigma_{los}$, and 
for a spiral galaxy its rotation velocity). \\
In the case of spiral galaxies, 
the depth of the
potential can be related to the circular velocity of the galaxy, which
 can be obtained either directly or from the apparent magnitude and the
redshift of the galaxy. Let us assume for simplicity that the
mass-to-light ratio of the galaxy is independent of its luminosity. Then
the typical scale
\begin{equation}
r_c \propto \sqrt{M} \ \ , \ {\rm that\  is,\  from\  Tully-Fisher:} \
r_c \propto \sqrt{L} \propto V_c^2 \ .
\end{equation}
\noindent Therefore, the photometry of the foreground galaxies  
can, to first approximation, scale their mass, from the 
calibrated Tully-Fisher relation,  as well as   
   their redshift, from a magnitude-redshift relation. The 
 main objective of the galaxy-galaxy lensing is then to calibrate 
the Tully-Fisher relation by measuring the physical 
scales $r_c^{*}$ and $L^{*}$ from the averaged polarization produced 
 by  the foreground galaxies onto the background lensed sources. \\
 The
 expected gravitational distortion is very weak: for foregrounds at
redshift $<z_l>=0.1$, backgrounds at $<z_s>=0.5$, and typical halos
with velocity dispersion of
 200 kms$^{-1}$ and  radius of 100 kpc, $\vert \gammag \vert \approx
$1\% at about 20 kpc from the center. But if the observations go
 to very faint magnitudes
there is a huge number of background lensed galaxies, so that the
weakness of the signal is compensated by the large statistics.
 \\
\begin{figure}
\caption{\label{brainerd.eps} {\it Angular variation of polarization produced by
weak lensing of foreground galaxies on the background (lensed) sources in the Brainerd
et al (1996) sample. The lines show theoretical expectations for three models
 of halos having different velocity dispersion and scales.
}
}
\end{figure}
The first reliable results came from deep  
 sub-arcsond seeing CCD observations (Brainerd et al 1996).
The distortion was compared with simulations, based on
analytical profiles assumed for the dark matter halos as well 
 as  the Tully-Fisher
relation and a magnitude-redshift relation,
   in order to relate mass models to observations.
 They detected a significant polarization
of about 1\% , averaged over separation between 5" and 34" (see Figure
 9).  They
concluded that halos smaller that 10$h^{-1}$ kpc are excluded at a
2$\sigma$ level, but the data are compatible with halos of size
larger that 100$h^{-1}$ kpc and circular velocities of 200 kms$^{-1}$.
\\
The HST data look perfectly suited for this kind of program which
demands high image quality and the observation of many field galaxies.
Griffiths et al (1996)
 used the Medium Deep Survey
 (MDS) and measured the
distortion produced by foreground elliptical and spiral galaxies. They
found similar results as Brainerd et al (1996)  but with a more
significant
signal for foreground elliptical than spiral galaxies. The comparison
 with shear signals expected from various analytical models seems  to
 rule out
 de Vaucouleur's law as mass density profile of ellipticals.
 Dell Antonio \& Tyson (1996) and Hudson et al (1998)  
analyzed the galaxy-galaxy lensing signal in the HDF. As
compared with the ground-based images or the MDS, the
field is small but the depth permits to use many background galaxies
 even on scale smaller than 5 arcseconds.  Furthermore, the UBRI
 data of the HDF permit to infer
 accurate photometric redshifts for the complete sample of galaxies.
 By comparing the lensing signal with predictions
from
an analytical model for the halo, Dell'Antonio \& Tyson 
 found  a significant distortion of about 7\% at 2" from the
halo center which corresponds to halos with typical circular velocities
of  less than 200 km.sec$^{-1}$.  All these results
  seem consistent with those
of  Brainerd et al.

\subsubsection{Galaxy-galaxy lensing on cluster galaxies}
As it has been shown from the observation of perturbations along giant arcs, 
 the clumpiness of dark matter on small-scale can be probed by anomalies in 
 lensing configurations (see Sect. 4.2.3).  
With the details visible on the
HST images of arclets in A2218, AC114 or A2390,
 the sample of halos which can be constrained by this
 method is much larger and can provide more
significant results by extending the method to perturbations along 
 arclets. The number of details permits also to
 use more sophisticated methods of investigation.
 \\
The simplest strategy is to
start with an analytical potential which reproduces the general
features of the shear pattern of HST images, and in a second step, to
include
in the model  analytical
halos around the brightest cluster members. In practice, additional mass
components
 are put in the model in order to interpret the arc(let)s which cannot
be
easily explained by the simple mass distribution.  Some guesses are done
in order to pair unexplained  multiple images. The colors of
the arc(let)s as well as their
morphology help a lot to make these associations.
This approach has been  proposed
 by Natarajan \& Kneib (1997), and Natarajan et al (1998). The detailed
study done
in AC114 by Natarajan et al (see Figure 10)
 indicates that about 10\% of the dark matter is associated with
 halos of cluster galaxies. These halos
have truncation radii smaller than field galaxies ($r_t \approx $15kpc)
with a general trend
of S0-galaxies to be even more truncated than the other galaxies.  If
this result is confirmed it would
be a direct evidence that truncation by tidal stripping is really
efficient in rich clusters of galaxies.  This result is somewhat
contradictory with the absence a clear decrease of rotation curves
 of spiral galaxies in nearby clusters (Amram et al 1993) which
 is interpreted as a proof that massive halos of galaxies still exist
 in cluster galaxies. However,
it could be explained if the spirals which have
been analyzed appear to be in the cluster center only by projection
effects
 but  are not really located
in the very dense region of the clusters where stripping is
 efficient. \\
\begin{figure}
\caption{\label{ac114.eps} {\it A nice example of a complete lensing analysis 
using arc(let)s. The  figure shows the mass density contours of AC114 
overlayed to the HST image and obtained from 
 a fitting of potential wells which include the clumps associated to the clusters 
and the halos of the galaxies. Therefore, both the massive clumps and the
perturbations are included. This permits to estimate the mass and size or dark 
halos of galaxies (from Natarajan \& Kneib 1997).   
}
}
\end{figure}
Geiger \& Schneider (1997, 1998) used a maximum likelihood
analysis which explores simultaneously the distortions induced by the
cluster as a whole and by its individual galaxies.  They applied this
analysis
 to the HST data of Cl0939+47 and reached  similar conclusions
 as Natarajan et al (1998).  Several issues limit the reliability of
their
analyses and of the other methods as well (Geiger \& Schneider 1998).
First, depending on the slope
of the mass profile of the cluster, the contributions of the cluster
 mass density and of the cluster galaxies  may be difficult to
 separate.  Second, it is necessary to have a realistic model for
 the redshift distribution of the background and foreground
galaxies.  Finally, the mass sheet degeneracy  is also
an additional
source of uncertainties.  Regarding these limitations, Geiger \&
Schneider discuss the capability of the
galaxy-galaxy lensing in clusters to provide valuable constrains
 on the galactic halos from  the data they have in hands.  Indeed,
  some of the issues they
raised can be solved, like for instance the redshift distribution
of the galaxies. It would be interesting to look into more details
 how the analysis could be improved with more and better data.

\section{Conclusion and future prospects}
The measurement of mass of galaxies and cluster of galaxies made a 
major step after the discovery of gravitational lenses like
 rings, arcs and arclets. \\
The results on clusters of galaxies seem quite robust. 
 In summary, the total masses and
 the  mass-to-light ratio recovered from lensing theory are similar to
those found by other techniques, which  confirms that investigation of dark matter 
in clusters favors $\Omega<0.4$. 
  The new result is the evidence that dark matter 
is strongly concentrated at the cluster center.
 More interesting, it seems that 
 the geometry of the light distribution of the brightest cluster
members traces the dark matter with a good accuracy. Furthermore, 
despite the discrepancies found between the X-ray mass and the lensing
mass on small scales, the agreement is good on Megaparcsec scales. This
 is still a matter of debates, but the dichotomies between large scale
and small scale as well as between non-cooling flows and cooling flow
clusters are strong arguments that the discrepancy is produced by over
simplifications on the dynamical stage of the gas.   \\
The investigation of galaxies are more debated. Though Einstein rings 
 provide very accurate mass, the galaxy-galaxy lensing approach is
still at its infancy and needs much more attention before providing 
something reliable. \\
In the future,  the investigation of gravitational systems will be
extended to nearby clusters. For those systems, the shear amplitude is
small but each angular scale probes a much smaller physical scale, so
that one can use much more galaxies than in distant systems in order
to map the same physical scale.   Groups of galaxies will be soon 
 analyzed extensively as well. In parallel, we are still looking for dark 
clusters that gravitational lensing could reveal easily. It these systems 
do exist, then their discovery would be a major breakthrough.
\\
Indeed, 
 the wide field CCD cameras and the future NGST
observations will permit to increase considerably  the scientific 
 return from lensing studies. The two key
technical points are the understanding of systematics and the
statistics. Both wide field CCD cameras and HST/NGST will provide important
insights and intrinsically much better data that present-day
observations. In parallel, since the masses are scaled by angular
distances, the redshifts of the lenses and the sources are needed. 
 The new giant telescopes equipped with visible and infrared
spectrographs, as well as the joint use of visible and near infrared 
photometric redshifts open a new area. We expect that the number of 
Einstein rings and giant arcs with redshift will increase significantly
 in the next decade,  
putting much better constrains on the mass and mass-to-light ratios 
of galaxies and clusters. 

\section*{Acknowledgments}
I am grateful to M. Lachi\`eze-Rey for his invitation
to give  this lecture in Carg\`ese. I thank F. Bernardeau, D. Elbaz, 
B. Fort, J.-P. Kneib, J. Rich, P. Schneider, L. van Waerbeke for the numerous stimulating 
discussions.

\end{document}